\documentstyle[prb,preprint,aps,epsfig]{revtex}
\begin{document} 
%\bibliographystyle{unsrt} 
%\draft
\title{Distribution of Return Intervals of Extreme Events} 

\author{Cecilia Pennetta}

\address{Dipartimento di Ingegneria dell'Innovazione,
Universit\`a di Lecce and CNR-INFM, \\National Nanotechnology Laboratory,
Via Arnesano, I-73100, Lecce, Italy,\\cecilia.pennetta@unile.it }

%\date{\today}

\maketitle

\begin{abstract}
The distribution of return intervals of extreme events is studied in time 
series characterized by finite-term correlations with non-exponential decay. 
Precisely, it has been analyzed the statistics of the return intervals of 
extreme values of the resistance fluctuations displayed by resistors with 
granular structure in nonequilibrium stationary states. The resistance 
fluctuations are calculated by Monte Carlo simulations using a resistor 
network approach. It has been found that for highly disordered networks, when 
the auto-correlation function displays a non-exponential and non-power-law 
decay, the distribution of return intervals of the extreme values is  
a stretched exponential, with exponent independent of the 
threshold. 
\end{abstract}  

\vspace{1.5cm}\hspace{1.0cm}

%{\small{\bf keywords}: Extreme values in random processes, 
%Fluctuation phenomena,Time series analysis} 

\narrowtext

\vspace{2.2cm}

\section{Introduction}  
Fluctuations of prices in financial markets, wind speed data or daily
precipitations in a given place for the same time windows are typically
described by time series $x(t)$ made by uncorrelated records 
\cite{storch,kantz_2005}. In these cases, by considering the return intervals 
$r_q$ of extreme events associated with the overcoming of a given threshold 
$q$ (i.e. the time intervals between two consecutive occurences of the 
condition $x(t)>q$), it can be seen that the $r_q$ are exponentially 
distributed \cite{storch,kotz,sornette}. In other terms, the probability 
density function (PDF) of the distribution of the $r_q$ is given by 
\cite{storch,kotz,sornette} : 
%
%\vspace{ -0.5truecm}
\begin{equation}
P_q(r)=(1/{R_q}) \exp (-r/{R_q})  \label{poisson} 
\end{equation}
%\vspace{ -0.5truecm}
%
where $R_q$ is the mean return interval. Of course, the higher the threshold 
$q$, the bigger is the value of $R_q$. On the other hand, in the last 
years it has become clear that several other important examples of 
time series display long-term correlations 
\cite{kantz_2005,bunde_physa2003,bunde_prl2005}. This is the case 
of physiological data (heartbeats \cite{bunde_prl2000,ashkenazy} and 
neuron spikes \cite{davidsen}), hydro-meteorological records (daily 
temperatures \cite{kantz_2005,bunde_physa2003,koscielny_prl98}), 
geophysical or astrophysical data (occurrence of earthquakes 
\cite{bak,corral} or solar flares \cite{boffetta}), internet traffic 
\cite{bunde_physa2003} and stock market volatility \cite{kantz_2005,liu} 
records. Long-term correlated series are characterized by an auto-correlation 
function, $C_x(s)$, decaying as a power-law: 
\begin{equation}
%\vspace*{-0.2cm}   
C_x(s) = <x_ix_{i+s}> = { 1 \over N-s} \sum_{i=1}^{N-s} x_ix_{i+s}
\sim s^{-\gamma_x}  \label{long_corr}  
%\vspace*{-0.2cm}  
\end{equation}  
with exponent $\gamma_x$ (correlation exponent) between 0 and 1. In this case,
the mean correlation time $\tau$, given by the integral over $s$ of $C_x(s)$,
diverges. The effect of long-terms correlations on the statistics of the 
$r_q$ has been first studied by Bunde et al. \cite{bunde_physa2003}
and the conclusions of this study were the following. i) The mean return 
interval $R_q$ is left unchanged by the presence of long-terms correlations. 
ii) The distribution of return intervals becomes a stretched exponential:
%
%\vspace{ -0.2truecm}
\begin{equation}
P_q(r) = a \ \exp\bigl[ -\bigl(b \ \ r / R_q \bigr)^{\gamma} \bigr]  \label{stretch}  
\end{equation}   
%\vspace{ -0.2truecm}
%
where the two exponents $\gamma$ and $\gamma_x$ were found to be equal.
iii) The return intervals themself are long-term correlated with a correlation
exponent $\gamma' \approx \gamma_x$. As noted by Altmann and Kantz
in a very recent paper \cite{kantz_2005}, the result i), obtained by 
Bunde et al. \cite{bunde_physa2003} by statistical arguments, can be 
identified with Kac's lemma \cite{kac} introduced in the context of 
dynamical systems. The results ii) and iii) were obtained in Ref. 
\cite{bunde_physa2003} on the ground of numerical calculations performed on 
long-term correlated and Gaussian time series generated by the algorithm
described in Ref. \cite{makse}. It must be noted that the result ii) only 
applies to linear time series (i.e. to series whose properties are completely 
defined by the power spectrum and by the probability distribution, regardless 
of the Fourier phases) \cite{kantz_2005,bunde_prl2005}. Apart from this 
restriction, the stretched exponential distribution of the $r_q$ seems to be a 
general feature in presence of long-term correlations in a time series 
\cite{kantz_2005,bunde_prl2005}. It must be underlined that this fact has 
important consequences on the observation of extreme events: indeed it 
implies a strong enhancement of the probability of having return intervals 
well below and well above $R_q$, in comparison with the occurrence of 
extreme events in an uncorrelated time series. Furthermore, it must be noted 
that the stretched exponential distribution depends on the parameter $\gamma$
only, being $a$ and $b$ in Eq.~(\ref{stretch}) functions of $\gamma$, 
as shown in Ref. \cite{kantz_2005}. 

In this paper it will be analyzed the effect on the distribution of 
the $r_q$ of the presence of finite-term correlations with non-exponential 
decay, a situation which can occur in systems which are approaching 
criticality, where intermediate behaviors, consisting in a non-exponential and 
non-power-law decay of correlations can emerge \cite{sornette}. 
Precisely, in the following section it will be studied the statistics of the 
return intervals of extreme values of the resistance fluctuations displayed by
a resistor with granular structure in a nonequilibrium stationary state 
\cite{pen_pre,pen_fnl,pen_physa,pen_ng_fn04,pen_prb}.

%\vspace{ -0.5truecm}
\section{Method and Results}  
%\vspace{ -0.5truecm}
The time series analyzed in this paper consist in the resistance fluctuations 
of a thin resistor with granular structure, in contact with a thermal bath at 
temperature $T_0$ and biased by an external current $I$. The resistance values
are calculated by using the stationary and biased resistor network (SBRN) 
model \cite{pen_pre,pen_fnl,pen_physa,pen_ng_fn04,pen_prb,pen_upon99}. 
This model describes a thin film with granular structure as a two-dimensional 
resistor network in a stationary state determined by the competition between 
two stochastic processes, breaking and recovery of the elementary resistors.
A broken elementary resistor (with resistance 10$^9$ higher than the 
resistance corresponding to a normal elementary resistor) can be associated
with a high resistivity region within the conducting material. 
Both processes are thermally activated and biased by the external current. The
resistance of the network and its fluctuations are calculated by Monte Carlo 
simulations \cite{pen_pre,pen_fnl,pen_prb}. Within this model, the level 
of intrinsic disorder in the network (average fraction of broken resistors 
in the vanishing current limit \cite{pen_prl_stat}) is controlled by a 
characteristic parameter \cite{pen_ng_fn04}: 
$\lambda \equiv (E_D -E_R)/k_B T_0$, where $E_D$ and $E_R$ are the activation 
energies respectively of the breaking and recovery processes. 
The intrinsic disorder parameter $\lambda$ ranges between: 
$\lambda_{min}<\lambda<\lambda_{max}$, where $\lambda_{max}$ corresponds to an
homogeneous resistor (perfect network) and 
$\lambda=\lambda_{min} \approx 0$ to the maximum level of intrinsic disorder 
compatible with a stationary state of the network (stationary resistance 
fluctuations) \cite{pen_fnl,pen_physa,pen_ng_fn04,pen_prb,pen_prl_stat}.

In addition to this intrinsic disorder, the SBRN model considers also the
presence of a disorder driven by the external current $I$. As a consequence, 
for a given value of $\lambda$, and for a network of given size, 
nonequilibrium stationary states of the network exist only when 
$I\leq I_B$ (breakdown threshold). For contrast, when $I>I_B$ the network 
undergoes an electrical breakdown associated with an irreversible divergence 
of its resistance \cite{pen_pre,pen_fnl,pen_physa,pen_prb}.  
For an arbitrary value of $\lambda$ this breakdown corresponds to a first 
order transition from a conducting to a non-conducting state of the network
\cite{pen_fnl,pen_physa,pen_ng_fn04}. However, at decreasing $\lambda$
values, when $\lambda  \rightarrow \lambda_{min}$, the system becomes more
and more close to its critical point \cite{pen_fnl,pen_physa,pen_ng_fn04}. 

All the details about the SBRN model and its results can be found
in Refs. \cite{pen_pre,pen_fnl,pen_physa,pen_ng_fn04,pen_prb}. However, 
it must be noted that this model provides a good description of many features
associated with nonequilibrium stationary states and with the electrical
instability of composites and granular materials 
\cite{pen_pre,pen_fnl,pen_physa,bardhan}, including the electromigration 
damage of metallic lines \cite{pen_prb}. Finally, it must be underlined that,
apart from the specific system described by the SBRN model, the method 
adopted here for generating the time series can be also viewed as a pure 
numerical algorithm for generating numerical series with 
different and tunable correlation properties.

\begin{figure}
{\begin{minipage}{7.5truecm}
\begin{center}
 \epsfig{figure=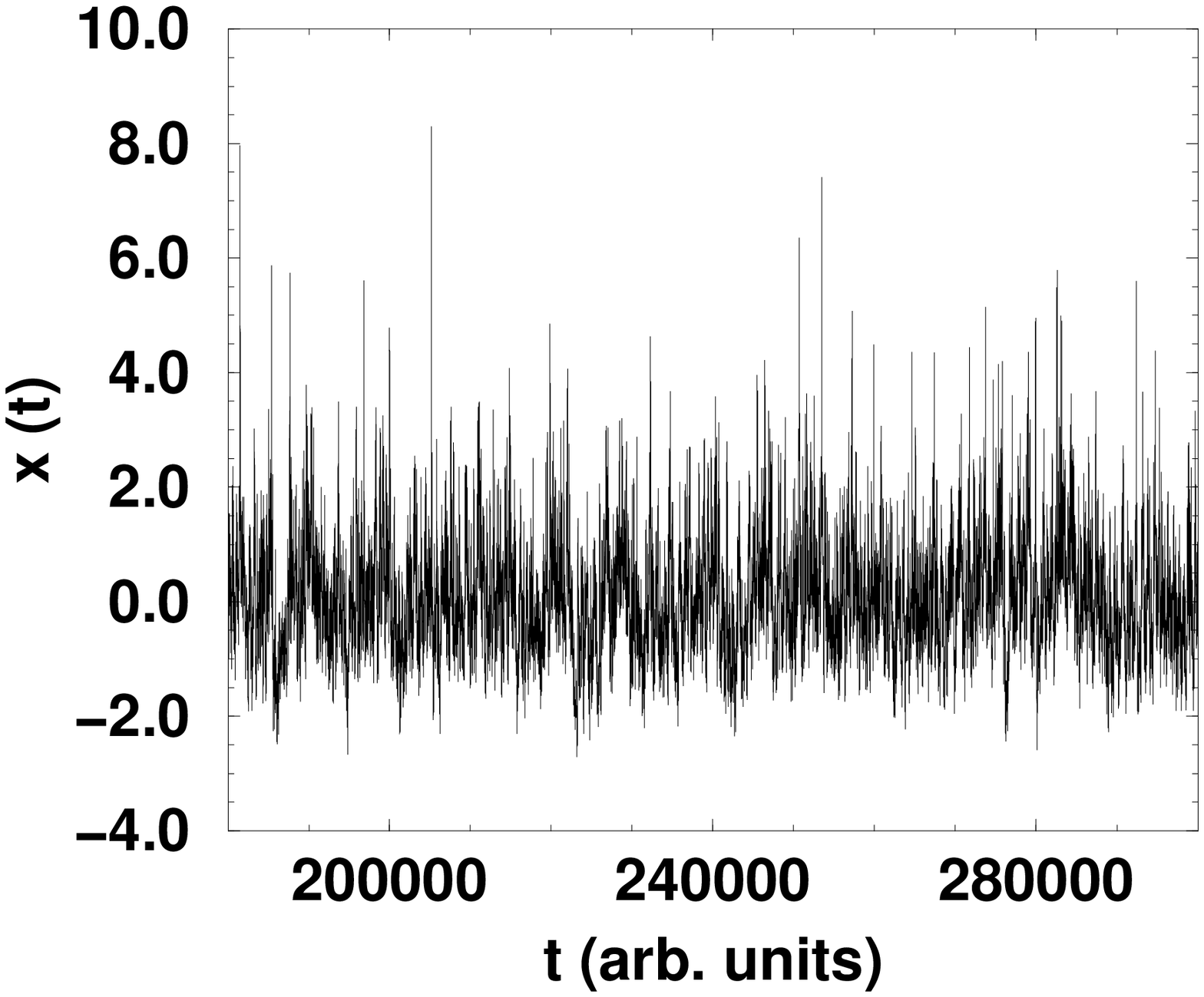,width=6.8cm}    
  \caption{Normalized resistance fluctuations versus time (this last is
   expressed in simulation steps). The resistance values have been normalized 
   to provide a zero average and a unit variance for $x(t)$.}
\end{center}
\label{fig:1}
\end{minipage}
\vskip -10.1truecm\hskip 8.0truecm
\begin{minipage}{7.5truecm}
\begin{center}
  \epsfig{figure=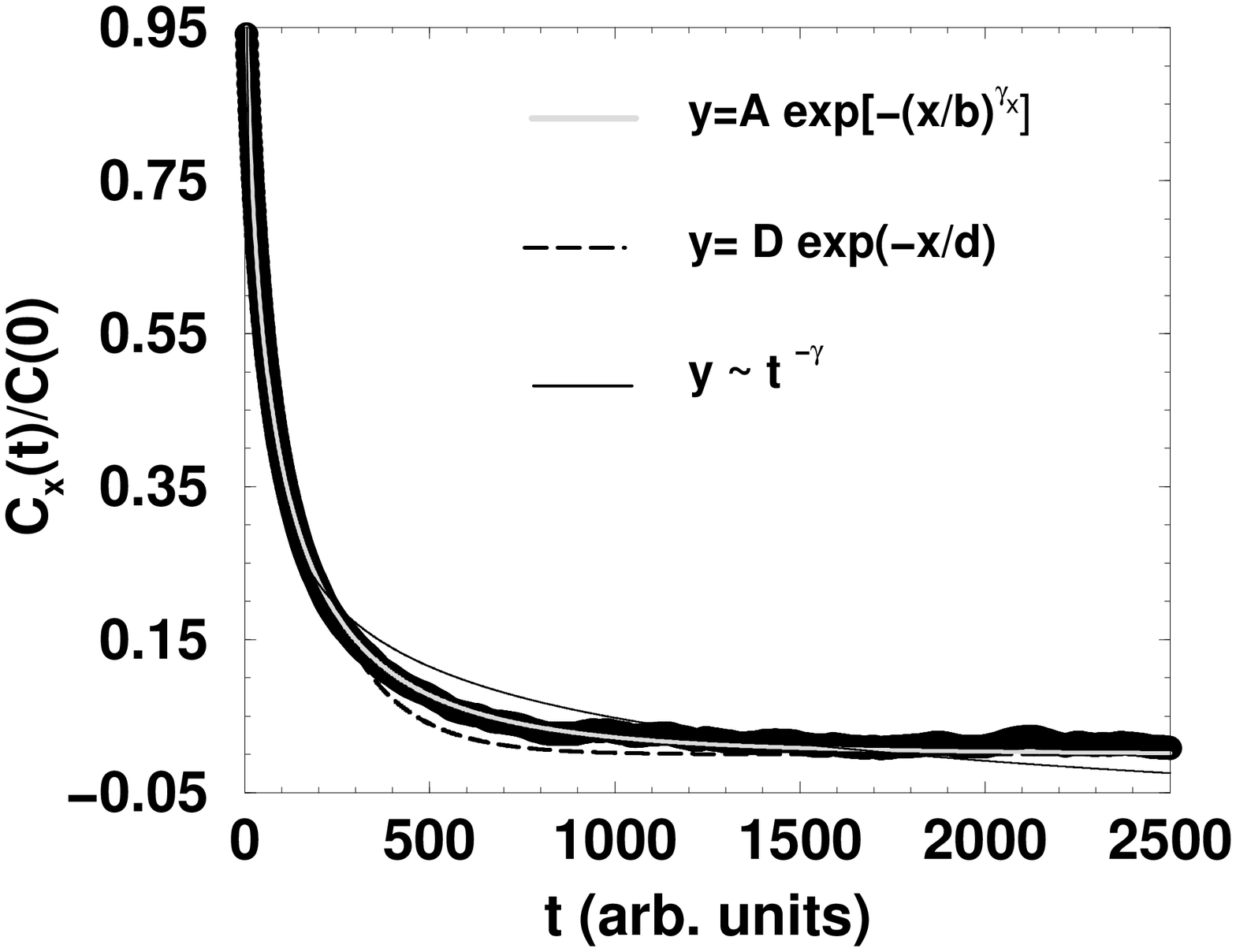,width=7.4cm}       
  \caption{Auto-correlation function of the time series in Fig. 1 
  (black-thick curve). The grey, black-dashed and black-solid curves represent 
  the best-fit with a stretched exponential, an exponential and a power-law, 
  respectively.}
\end{center}
\label{fig:2}
\end{minipage}
}
\end{figure}

Then, by indicating with $R$ the resistance of the network (expressed in 
$\Omega$) and with $t$ the time (expressed in iteration steps), long $R(t)$ 
time series (typically made of $1\div 2 \times 10^6$ records) have been 
generated and analyzed for different values of $\lambda$, of the
external current and of the network size. Precisely, normalized series
with zero average and unit variance have been considered: 
$x(t) \equiv (R(t)-<R>)/\sigma$, where $<R>$ is the average value of the 
network resistance and $\sigma$ the root-mean-square deviation from the 
average. The analysis has been performed by calculating the auto-correlation 
function and the PDF of the $x$ records, the return intervals $r_q$ of 
the extreme values for different threshold $q$ and their distribution 
$P_q(r)$ (the values of $q$ are expressed in units of $\sigma$). For small
$\lambda$ values (high level of intrinsic disorder), it has been found that
$C_x$  displays a non-exponential and non-power-law decay. 
This behavior is different from that obtained for high $\lambda$ values
(low level of intrinsic disorder), where $C_x$ decays exponentially 
(consistently with the Lorentzian power spectrum reported in Refs. 
\cite{pen_pre,pen_fnl,pen_ng_fn04}). By focusing on this situation, of 
interest for systems which are approaching criticality, in the following 
of this section results will be shown concerning a network of size 
$125 \times 125$, biased by a current $I=I_B=0.011$ A and obtained by taking 
$\lambda=0.33$.

\begin{figure}
{\begin{minipage}{7.5truecm}
\vskip 0.5truecm
 \begin{center} 
 \epsfig{figure=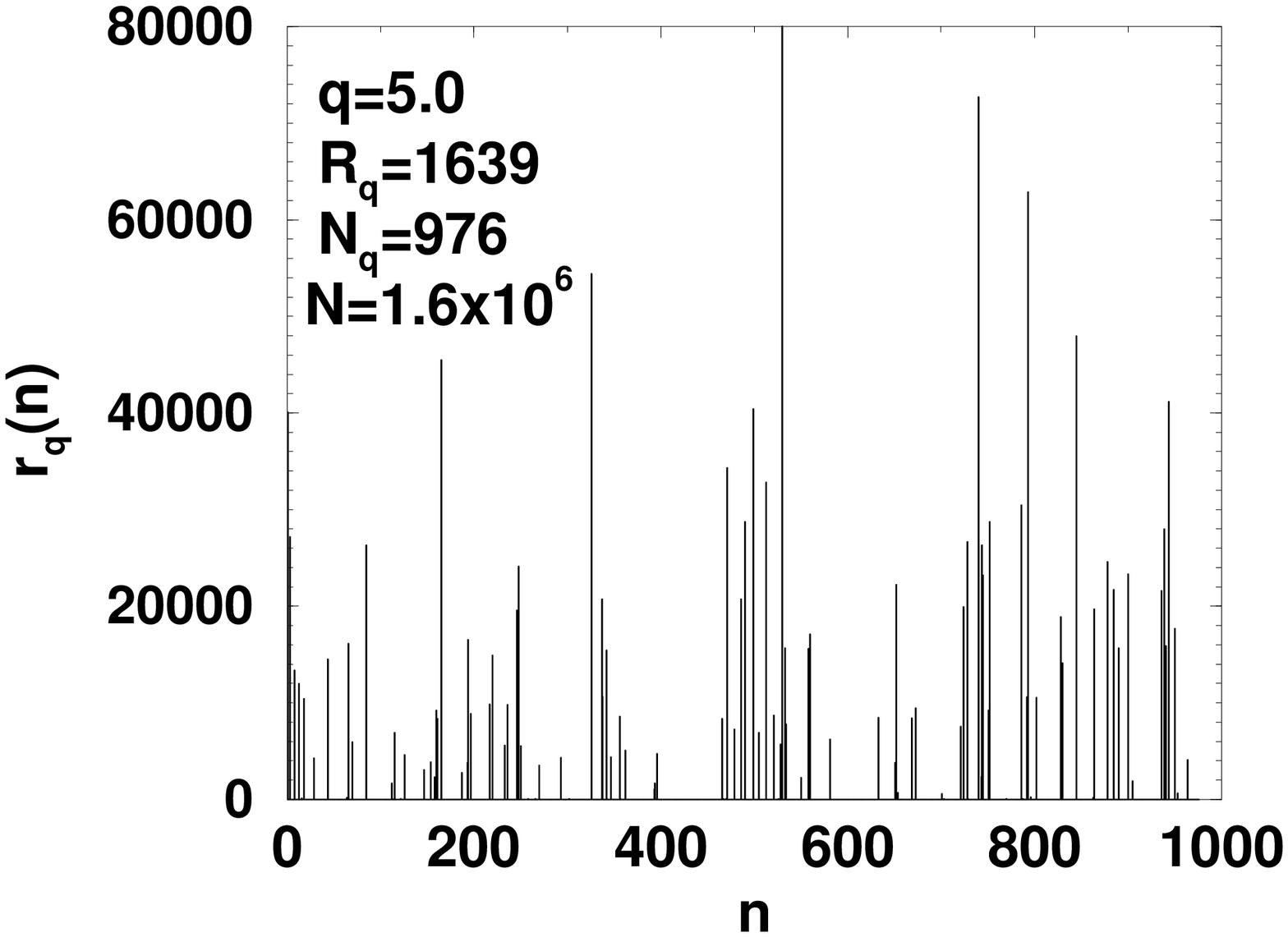,width=7.4cm}    
 \caption{Return intervals of extreme values above the threshold 
 $q=5$ (in units of root-mean square deviation) for the time series of 
 Fig. 1.}
\end{center}
\label{fig:3}
\end{minipage}
\vskip -9.0truecm\hskip 8.0truecm
\begin{minipage}{7.5truecm}
\begin{center}
  \epsfig{figure=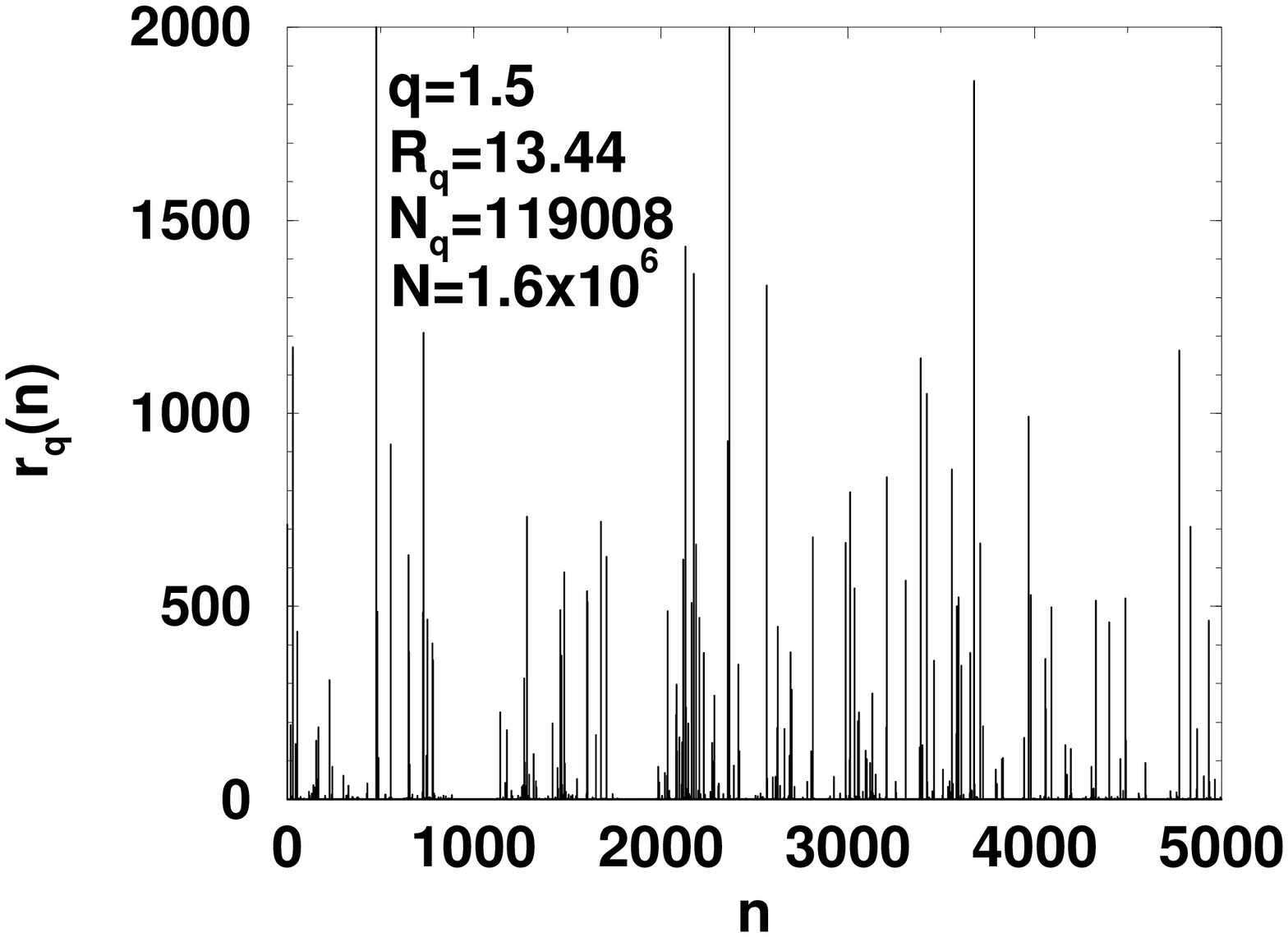,width=7.4cm}  
  \caption{Return intervals of extreme values above the threshold 
  $q=1.5$ for the same time series of the previous figures. Here 
  only the first 5000 intervals are shown.}   
\end{center}
\label{fig:4}
\end{minipage}
}
\end{figure}
 
The $x(t)$ time series is shown in Fig. 1 (only a small portion of the total
number of records, $N=1.6 \times 10^6$, is reported in this figure). As evident
from Fig. 1, the resistance fluctuations exhibit a strong non-Gaussianity
and actually their PDF is well described by the 
Bramwell-Holdworth-Pinton distribution \cite{bramwell_nat}, as discussed in 
Refs. \cite{pen_physa,pen_ng_fn04}. The auto-correlation function of $x(t)$ 
is reported in Fig. 2. The function significantly deviates from a single 
exponential and from a power-law while it is well fitted by a stretched 
exponential:
\begin{equation}
C_x(s)=A\ $ exp$[(-s/b)^\gamma)]  \label{stretch_cx}  
\end{equation}   
with the following values of the fitting parameters: $A=1.23$, $b=74.9$ and 
$\gamma=0.54$. Many other functions have also been considered for the best-fit
of the $C_x$ data. However, it has been found that the stretched exponential 
optimizes the best-fit procedure with the minimum numbers of fitting 
parameters. It must be remarked that a stretched exponential describes a 
behavior intermediate between a simple exponential decay (which is obtained 
for $\gamma=1$) and a constant behavior (a limit of power-law) for 
$\gamma \rightarrow 0$. Moreover, the correlation time corresponding to the 
expression (\ref{stretch_cx}) of $C_x$ is finite.

The sequence of the $N_q$ return intervals of the values above the 
threshold $q=5$ is plotted in Fig. 3 as a sequence of $N_q$ impulses. The 
figure shows that a succession of very short return intervals ($r_q \ll 100$,
the apparently empty portions of the horizontal axis) is followed by a 
succession of long intervals, indicating a strong clustering of the extreme 
events, a feature similar to that exhibited by the data of Bunde et al. 
\cite{bunde_prl2005} which instead concern long-term correlated records
and very different from what observed for uncorrelated time series 
\cite{upon05}. Thus, the clustering of the extreme events is present even if 
the $x$ records are not long-term correlated while they are characterized by a
finite correlation time. This clustering of the extreme values persists also 
by lowering the threshold. This is shown in Fig. 4 which reports the sequence 
of the return intervals obtained for $q=1.5$.

\begin{figure}  
\vskip 0.3truecm
{\begin{minipage}{7.5truecm}
\begin{center}
 \epsfig{figure=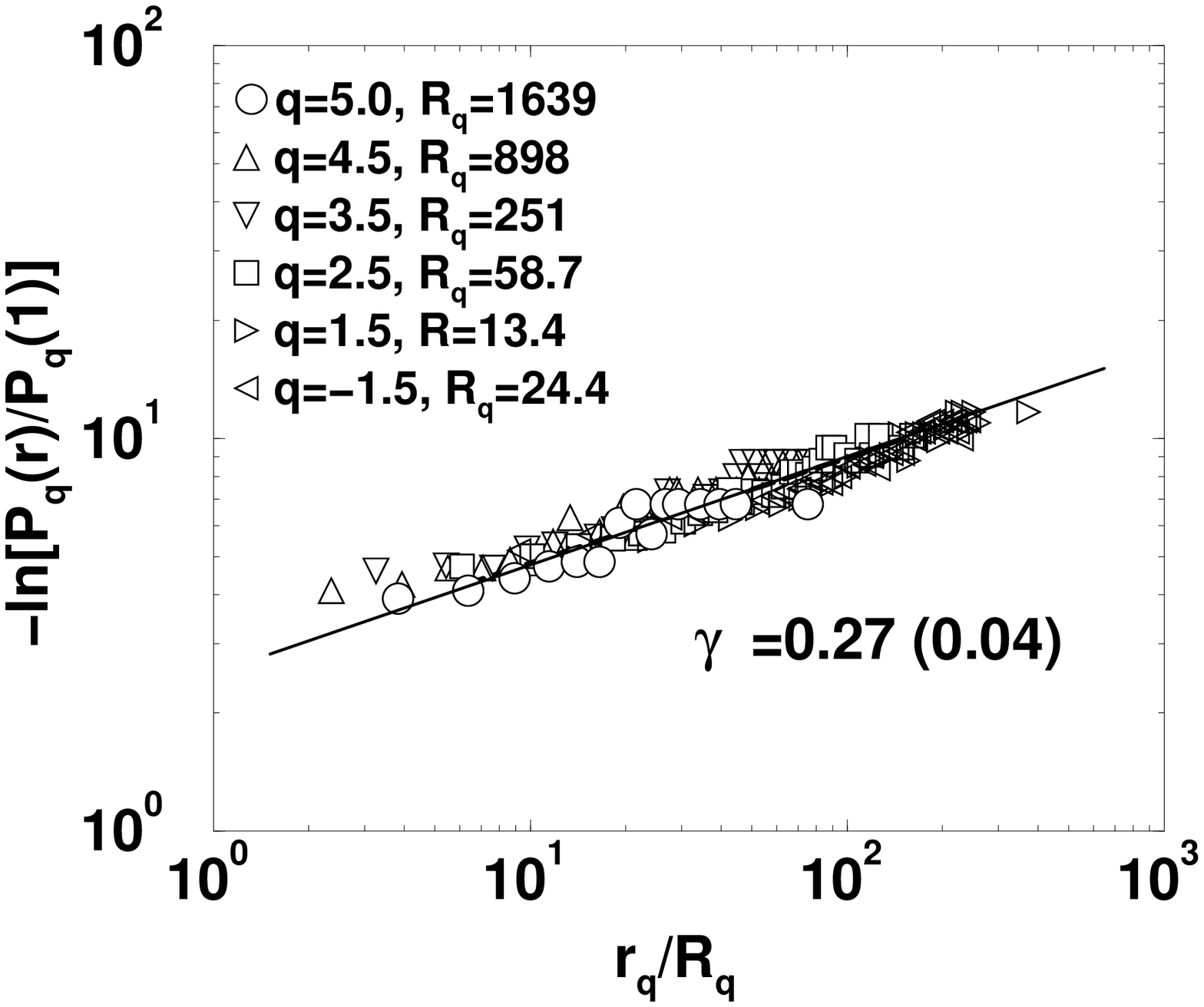,width=7.4cm}  
 \caption{Double-logarithmic plot of the normalized probability density 
  of the return intervals for different thresholds $q$.}
\end{center}
\label{fig:5}
\end{minipage}
\vskip -8.7truecm\hskip 8.0truecm
\begin{minipage}{7.5truecm}
\begin{center}  
 \epsfig{figure=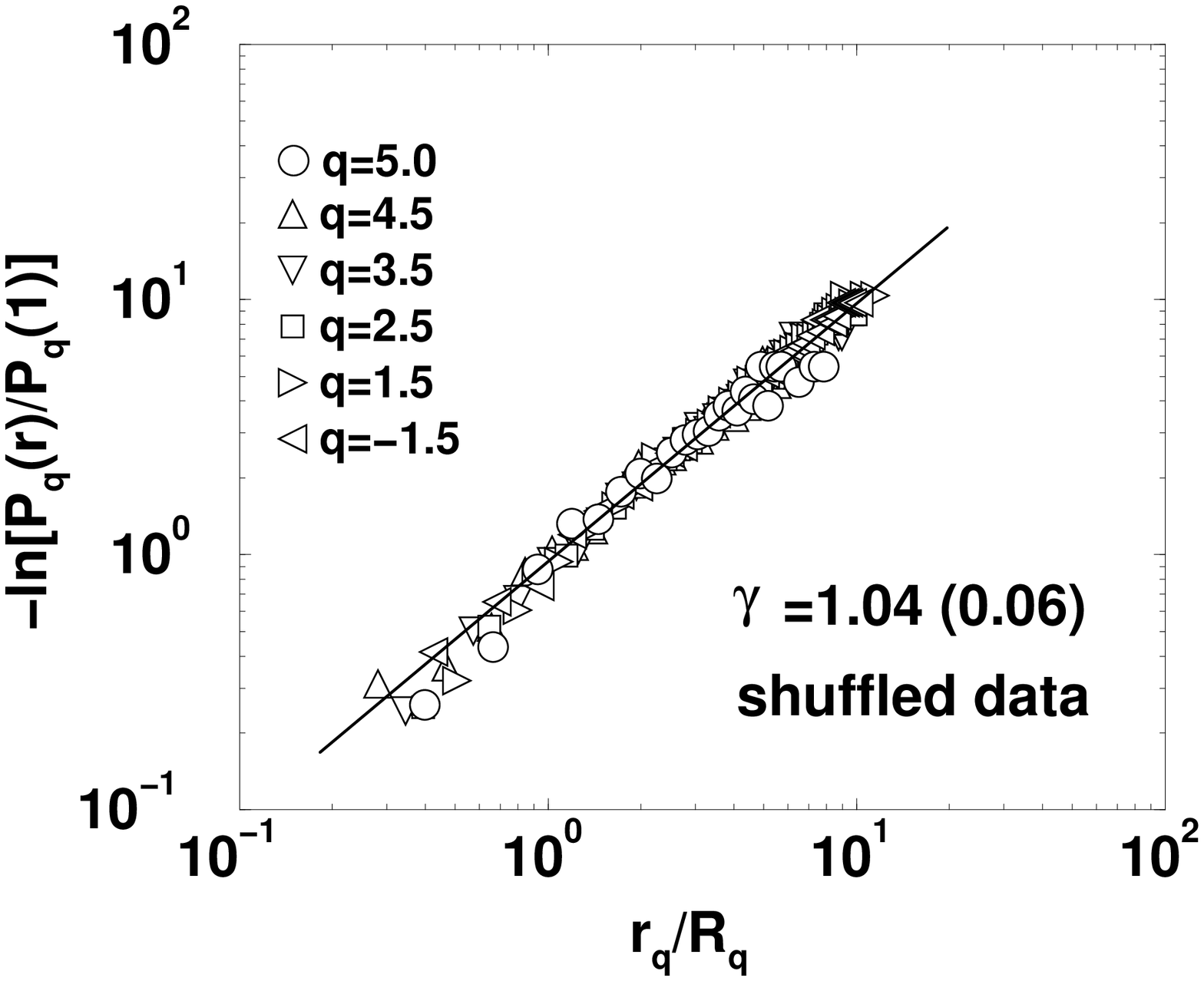,width=7.4cm}  
 \caption{The same as Fig. 5 but calculated after random shuffling the 
  x-records.}
\end{center}
\label{fig:6}   
\end{minipage}
}
\vskip 0.8truecm
\end{figure}

Figure 5 displays the probability density function $P_q(r)$ of the distribution
of the return intervals as a function of $r_q/R_q$  for different thresholds 
$q$ ranging from $-1.5$ to $5.0$ (the probability density has been normalized 
to $P_q(1) \ )$. A double-logarithmic plot of $P_q$ has been adopted because
in this representation a stretched exponential function with exponent $\gamma$
appears as a straight line of slope $\gamma$. Therefore Fig. 5 shows that the 
distribution of the return intervals of extreme values of the x-series is well
described by a stretched exponential and that the value of the exponent 
$\gamma$ is independent of the threshold $q$ in a large range of $q$-values. 
This occurs even in absence of long-term corrrelations and in presence of a
finite correlation time. This not-obvious result agrees with the conclusions 
of Altmann and Kantz formulated in their recent paper \cite{kantz_2005}. 
For comparison, Fig. 6 reports the probability density function $P_q(r)$ of 
the distribution of the return intervals obtained after random shuffling the 
records of the same x-series: in this case $\gamma=1$, i.e. the distribution 
of the $r_q$ is exponential, as it must be for uncorrelated time series. 

\section{Conclusions}  
The distribution of return intervals of extreme events has been 
studied in time series with finite-term correlations. Precisely, it has been 
analyzed the distribution of return intervals of extreme values of the 
resistance fluctuations displayed by a resistor with granular structure in 
nonequilibrium stationary states. The resistance fluctuations were calculated 
by using the SBRN model based on a resistor network approach 
\cite{pen_pre,pen_fnl,pen_physa,pen_ng_fn04,pen_prb}. It has been found that 
for highly disordered networks, when the auto-correlation function displays a 
non-exponential and a non-power-law decay, the distribution of the $r_q$ is
well described by a stretched exponential with exponent $\gamma$ largely 
independent of the threshold $q$. This result shows that the stretched 
exponential distribution describes the distribution of the return 
intervals of extreme events not only when long-term correlations are present
in the time series \cite{kantz_2005,bunde_physa2003,bunde_prl2005}, but also 
when finite-term correlations exist among the records, characterized by a
non-exponential decay, a situation typical of systems which are approaching
criticality. 

\vspace*{0.5cm}  
%\section*{ACKNOWLEDGMENTS}
Partial support from SPOT NOSED project IST-2001-38899 of E.C. and from 
MIUR cofin-03 project "Modelli e misure in nanostrutture" is acknowledged. 
The author thanks S. Ruffo (University of Florence, Italy), P. Olla 
(ISAC-CNR, Lecce, Italy), G. Salvadori and E. Alfinito (University of Lecce, 
Italy) for helpful discussions.

\end{document}